\documentclass[conference]{IEEEtran}
\ifCLASSINFOpdf
\else
\fi
\begin{document}
%
% paper title
% Titles are generally capitalized except for words such as a, an, and, as,
% at, but, by, for, in, nor, of, on, or, the, to and up, which are usually
% not capitalized unless they are the first or last word of the title.
% Linebreaks \\ can be used within to get better formatting as desired.
% Do not put math or special symbols in the title.
\title{Nonsense Attacks on Google Assistant}

% author names and affiliations
% use a multiple column layout for up to three different
% affiliations
\author{\IEEEauthorblockN{Mary K. Bispham}
\IEEEauthorblockA{Department of Computer Science\\
University of Oxford\\
Email: mary.bispham@cybersecurity.ox.ac.uk}
\and
\IEEEauthorblockN{Ioannis Agrafiotis}
\IEEEauthorblockA{Department of Computer Science\\
University of Oxford}
\and
\IEEEauthorblockN{Michael Goldsmith}
\IEEEauthorblockA{Department of Computer Science\\
University of Oxford}}

% conference papers do not typically use \thanks and this command
% is locked out in conference mode. If really needed, such as for
% the acknowledgment of grants, issue a \IEEEoverridecommandlockouts
% after \documentclass

% for over three affiliations, or if they all won't fit within the width
% of the page, use this alternative format:
% 
%\author{\IEEEauthorblockN{Michael Shell\IEEEauthorrefmark{1},
%Homer Simpson\IEEEauthorrefmark{2},
%James Kirk\IEEEauthorrefmark{3}, 
%Montgomery Scott\IEEEauthorrefmark{3} and
%Eldon Tyrell\IEEEauthorrefmark{4}}
%\IEEEauthorblockA{\IEEEauthorrefmark{1}School of Electrical and Computer Engineering\\
%Georgia Institute of Technology,
%Atlanta, Georgia 30332--0250\\ Email: see http://www.michaelshell.org/contact.html}
%\IEEEauthorblockA{\IEEEauthorrefmark{2}Twentieth Century Fox, Springfield, USA\\
%Email: homer@thesimpsons.com}
%\IEEEauthorblockA{\IEEEauthorrefmark{3}Starfleet Academy, San Francisco, California 96678-2391\\
%Telephone: (800) 555--1212, Fax: (888) 555--1212}
%\IEEEauthorblockA{\IEEEauthorrefmark{4}Tyrell Inc., 123 Replicant Street, Los Angeles, California 90210--4321}}

% use for special paper notices
%\IEEEspecialpapernotice{(Invited Paper)}

% make the title area
\maketitle

% As a general rule, do not put math, special symbols or citations
% in the abstract
\begin{abstract}
This paper presents a novel attack on voice-controlled digital assistants using nonsensical word sequences. We present the results of experimental work which demonstrates that it is possible for malicious actors to gain covert access to a voice-controlled system by hiding commands in apparently nonsensical sounds of which the meaning is opaque to humans. Several instances of nonsensical word sequences were identified which triggered a target command in a voice-controlled digital assistant, but which were incomprehensible to humans, as shown in tests with human experimental subjects. Our work confirms the potential for hiding malicious voice commands to voice-controlled digital assistants or other speech-controlled devices in speech sounds which are perceived by humans as nonsensical.
\end{abstract}

% no keywords

% For peer review papers, you can put extra information on the cover
% page as needed:
% \ifCLASSOPTIONpeerreview
% \begin{center} \bfseries EDICS Category: 3-BBND \end{center}
% \fi
%
% For peerreview papers, this IEEEtran command inserts a page break and
% creates the second title. It will be ignored for other modes.
\IEEEpeerreviewmaketitle

\section{Introduction}
 
The growing popularity of voice-controlled digital assistants, such as Google Assistant, brings with it new types of security challenges. Input to a speech interface is difficult to control. Furthermore, attacks via a speech interface are not limited to voice commands which are detectable by human users. Malicious input may also come from the space of sounds to which humans allocate a different meaning to the system, or no meaning at all. In this paper, we show that it is possible to hide malicious voice commands to the voice-controlled digital assistant Google Assistant in `nonsense' words which are perceived as meaningless by humans. 

The remainder of the paper is structured as follows. Section II outlines the prior work in this area. Section III provides some relevant background on phonetics and speech recognition. Section IV details the methodology that was applied in the experimental work. This includes the process used to generate potential adversarial commands consisting of nonsensical word sounds. It also includes the processes for testing the response of Google Assistant to the potential adversarial commands, and for testing the comprehensibility of adversarial commands by humans. Section V presents the results of the experimental work. Section VI discusses the implications of the experimental results. Section VII makes some suggestions for future work and concludes the paper.  

\section{Prior Work}

The idea of attacking voice-controlled digital assistants by hiding voice commands in sound which is meaningless or imperceptible to humans has been investigated in prior work. Carlini et al. \cite{carlini2016hidden} have presented results showing it is possible to hide malicious commands to voice-controlled digital assistants in apparently meaningless noise, whereas Zhang et al.\cite{zhang2017dolphinatack} have shown that it is possible to hide commands in sound which is inaudible. Whereas this prior work demonstrated voice attacks which are perceived by humans as noise and silence, the aim of this work was to develop a novel attack based on `nonsense' sounds which have some phonetic similarity with the words of a relevant target command. In related work by Papernot et al. \cite{papernot2016crafting}, it was shown that a sentiment analysis method could be misled by input which was `nonsensical' at the sentence level, i.e. the input consisted of a nonsensical concatenation of real words. By contrast, this work examines whether voice-controlled digital assistants can be misled by input which consists of nonsensical word sounds. Nonsense attacks are one of the categories of possible attacks via the speech interface which have been identified in a taxonomy developed by Bispham et al. \cite{bispham2018taxonomy}. 

Outside the context of attacks via the speech interface, differences between human and machine abilities to recognise nonsense syllables have been studied for example by Lippmann et al. \cite{lippmann1997speech} and Scharenborg and Cooke \cite{scharenborg2008comparing}. Bailey and Hahn \cite{bailey2005phoneme} examine the relationship  between theoretical measures of phoneme similarity based on phonological features, such as might be used in automatic speech recognition, and empirically determined measures of phoneme confusability based on  human perception tests. Machine speech recognition has reached parity with human abilities in terms of the ability correctly to transcribe meaningful speech (see Xiong et al. \cite{xiong2016achieving}), but not in terms of the ability to distinguish meaningful from meaningless sounds. The inability of machines to identify nonsense sounds as meaningless is exploited for security purposes by Meutzner et al. \cite{meutzner2015constructing}, who have developed a CAPTCHA based on the insertion of random nonsense sounds in audio. The opposite scenario, i.e. the possible security problems associated with machine inability to distinguish sense from nonsense, has to the best of our knowledge not been exploited in prior work.

\section{Background}

The idea for this work was inspired by the use of nonsense words to teach phonics to primary school children.\footnote{See The Telegraph, 1st May 2014, ``Infants taught to read `nonsense words' in English lessons"} `Nonsense' is defined in this context as sounds which are composed of the sound units which are used in a given language, but to which no meaning is allocated within the current usage of that language. Such sound units are known as `phonemes'.\footnote{See for example https://www.britannica.com/topic/phoneme} English has around 44 phonemes.\footnote{See for example https://www.dyslexia-reading-well.com/44-phonemes-in-english.html} The line between phoneme combinations which carry meaning within a language and phoneme combinations which are meaningless is subject to change over time and place, as new words evolve and old words fall out of use (see Nowak and Krakauer \cite{nowak1999evolution}). The space of meaningful word sounds within a language at a given point in time is generally confirmed by the inclusion of words in a generally established reference work, such as, in the case of English, the Oxford English Dictionary.\footnote{See for example https://blog.oxforddictionaries.com/press-releases/new-words-added-oxforddictionaries-com-august-2014/} In this work, we tested the response of Google Assistant to English word sounds which were outside this space of meaningful word sounds, but which had a `rhyming' relationship with meaningful words recognised as commands by Google Assistant. The term `rhyme' is used to refer to a number of different sound relationships between words (see for example McCurdy et al. \cite{mccurdy2015rhymedesign}), but it is most commonly used to refer to a correspondence of word endings.\footnote{see https://en.oxforddictionaries.com/definition/rhyme} For the purposes of our experimental work we define rhyme according to this commonly understood sense as words which share the same ending. 

There are a number of features of speech recognition in voice-controlled digital assistants which might affect the processing of nonsense syllables by such systems. One of these features is the word space which the assistant has been trained to recognise. The number of words which a voice assistant such as Google Assistant can transcribe is much larger than the number of words which it can `understand' in the sense of being able to map them to an executable command. In order to be able to perform tasks such as web searches by voice and note taking, a voice-controlled digital assistant must be able to transcribe all words in current usage within a language. It can therefore be assumed that the speech recognition functionality in Google Assistant must have access to a phonetic dictionary of all English words. We conducted some preliminary tests to determine whether this phonetic dictionary also includes nonsense words, so as to enable the assistant to recognise such words as meaningless. Using the example of the nonsense word sequence `voo terg spron', we tested the response of Google Assistant to nonsense syllables by speaking them in natural voice to a microphone three times. The nonsense word sequence was variably transcribed as `bedtime song', `who text Rob', and `blue tux prom', i.e. the Assistant sought to match the nonsense syllables to meaningful words, rather than recognising them as meaningless. This confirmed the viability of our experiment in which we sought to engineer the matching of nonsense words to a target command. 

Another feature of speech recognition in voice assistant which might affect the processing of nonsense syllables is the influence of a language model. Modern speech recognition technology includes both an acoustic modelling and a language modelling component. The acoustic modelling component computes the likelihood of the acoustic features within a segment of speech having been produced by a given word. The language modelling component calculates the probability of one word following another word or words within an utterance. The acoustic model is typically based on Gaussian Mixture Models or deep neural networks (DNNs), whereas the language model is typically based on n-grams or recurrent neural networks (RNNs). Google's speech recognition technology as incorporated in Google Assistant is based on neural networks.\footnote{See Google AI blog, 11th August 2015, `The neural networks behind Google Voice transcription' https://ai.googleblog.com/2015/08/the-neural-networks-behind-google-voice.html} The words most likely to have produced a sequence of speech sounds are determined by calculation of the product of the acoustic model and the language model outputs. The language model is intended to complement the acoustic model, in the sense that it may correct `errors' on the part of the acoustic model in matching a set of acoustic features to words which are not linguistically valid in the context of the preceding words. This assumption of complementary functionality is valid in a cooperative context, where a user interacts via a speech interface in meaningful language. However, the assumption of complementarity is not valid in an adversarial context, where an attacker is seeking to engineer a mismatch between a set of speech sounds as perceived by a human, such as the nonsensical speech sounds generated here, and their transcription by a speech-controlled device. In an adversarial context such as that investigated here, the language model may in fact operate in the attacker's favour, in that if one `nonsense' word in an adversarial command is misrecognised as a target command word, subsequent words in the adversarial command will be more likely to be misrecognised as target command words in turn, as the language model trained to recognise legitimate commands will allocate a high probability to the target command words which follow the initial one. Human speech processing also uses an internal `lexicon' to match speech sounds to words (see for example Roberts et al. \cite{roberts2013aligning}). However, as mentioned above, unlike machines, humans also have an ability to recognise speech sounds as nonsensical. This discrepancy between machine and human processing of word sounds was the basis of our attack methodology for hiding malicious commands to voice assistants in nonsense words.

\section{Methodology}
The experimental work comprised three stages. The first stage involved generating from a set of target commands a set of potential adversarial commands consisting of nonsensical word sequences. These potential adversarial commands were generated using a mangling process which involved replacing consonant phonemes in target command words to create a rhyming word sound, and then determining whether the resulting rhyming word sound was a meaningful word in English or a `nonsense word'. For the purposes of this work, the Unix word list was considered representative of the current space of meaningful sounds in English. Word sounds identified as nonsense words were used to create potential adversarial commands. Audio versions of these potential adversarial commands were created using speech synthesis technology. The second stage of the experimental work was to test the response of the target system to the potential adversarial commands. The target system for experiments on machine perception of nonsensical word sequences was the voice-controlled digital assistant Google Assistant. The Google Assistant system was accessed via the Google Assistant Software Development Kit (SDK).\footnote{See https://developers.google.com/assistant/sdk/} The third stage of the experimental work was to test the human comprehensibility of adversarial commands which were successful in triggering a target action in the target system. 
\subsection{Adversarial Command Generation}

A voice-controlled digital assistant such as Google Assistant typically performs three generic types of action, namely information extraction, control of a cyber-physical action, and data input. The data input category may overlap with the control of cyber-physical action category where a particular device setting needs to be specified, eg. light color or thermostat temperature. The three generic action categories are reflected in three different command structures for commands to Google Assistant and other voice-controlled digital assistants. The three command structures are: vocative + interrogative (eg. `Ok Google, what is my IP address'), vocative + imperative (eg. `Ok Google, turn on the light'), and vocative + imperative + data (eg. 'Ok Google, take a note that cats are great'). For our experimental work, we chose 5 three-word target commands corresponding to 5 target actions, covering all three possible target action categories. These target commands were: ``What's my name" (target action: retrieve username, action category: information extraction), ``Turn on light" (target action: turn light on, action category: control of cyber-physical action), ``Turn off light" (target action: turn light off, action category: control of cyber-physical action), ``Turn light red" (target action: turn light to red, action category: data input), ``Turn light blue" (target action: turn light to blue, action category: data input). We originally included a sixth target command, which would have represented a second target command for the information extraction category: ``Who am I". However, no successful adversarial commands could be generated from this target command.  

A set of potential adversarial commands was created from the target commands using a mangling process. This mangling process was based on replacing consonant phonemes in the target command words to generate nonsensical word sounds which rhymed with the original target command word.\footnote{Our approach was inspired by an educational game in which a set of nonsense words is generated by spinning lettered wooden cubes - see https://rainydaymum.co.uk/spin-a-word-real-vs-nonsense-words/}
The target commands were first translated to a phonetic representation in the Kirschenbaum phonetic alphabet\footnote{See http://espeak.sourceforge.net/phonemes.html} using the `espeak' functionality in Linux. The starting consonant phonemes of each word of the target command were then replaced with a different starting consonant phoneme, using a Python script and referring to a list of starting consonants and consonant blends.\footnote{See https://k-3teacherresources.com/teaching-resource/printable-phonics-charts/} Where the target command word began with a vowel phoneme, a starting consonant phoneme was prefixed to the vowel. The resulting word sounds were checked for presence in a phonetic representation of the Unix word list, also generated with espeak, to ascertain whether the word sound represented a meaningful English word or not. If the sound did correspond to a meaningful word, it was discarded. This process thus generated from each target command a number of rhyming nonsensical phoneme sequences to which no English meaning was attached. Audio versions of the phoneme sequences were then created using espeak. A similar process was followed to generate a set of potential adversarial commands from the wake-up word `Hey Google'. In addition to replacing the starting consonants `H' and `G', the second `g' in `Google' was also replaced with one of the consonants which are found in combination with the `-le' ending in English.\footnote{See https://howtospell.co.uk/}  

Nonsensical word sequences generated from the `Hey Google' wake-up word and nonsensical word sequences generated from target commands which were successful respectively in activating the assistant and triggering a target action in audio file input tests (see Results section for details) were combined with one another to generate a set of potential adversarial commands for over-the-air tests. This resulted in a total of 225 nonsensical word sequences representing a concatenation of each of 15 nonsensical word sequences generated from the wake-up word with each of 15 nonsensical word sequences generated from a target command. Audio versions of these 225 nonsensical word sequences were generated using the Amazon Polly speech synthesis service, generating a set of .wav files.\footnote{See https://aws.amazon.com/polly/} Amazon Polly is the speech synthesis technology used by Amazon Alexa, hence the over-the-air tests represented a potential attack on Google Assistant with `Alexa's' voice. The audio contained a brief pause between the wake-up word and the command, as is usual in natural spoken commands to voice assistants. As Amazon Polly uses the x-sampa phonetic alphabet rather than the Kirschenbaum format, it was necessary prior to synthesis to translate the phonetic representations of the potential adversarial commands from Kirschenbaum to x-sampa format.

\subsection{Assistant Response Tests}
 The Google Assistant SDK was integrated in a Ubuntu virtual machine (version 18.04). The Assistant was integrated in the virtual machine using two options; firstly, the Google Assistant Service, and secondly the Google Assistant Library. The Google Assistant Service is activated via keyboard stroke and thus does not require a wake-up word, and voice commands can be inputted as audio files as well as over the air via a microphone. The Google Assistant Library, on the other hand, does require a wake-up word for activation, and receives commands via a microphone only. The Google Assistant Service could therefore be used to test adversarial commands for target commands and for the wake-up word separately and via audio file input rather than via a microphone. The Google Assistant Library could be used to test the activation of the Assistant and the triggering of a target command by an adversarial command in combination over the air, representing a more realistic attack scenario. 

Some additions to the source code for the Google Assistant Service were made in order to print the Assistant's spoken responses to commands to the terminal in text form, as well to print a confirmation of two non-verbal actions by the Assistant, namely `turn light red' and `turn light blue'. Similar amendments were made to the source code for the Google Assistant Library in order to print a confirmation of these two non-verbal actions for which the Assistant did not print a confirmation to the terminal by default.

We first tested the Assistant's response to plain-speech versions of each target command to confirm that these triggered the relevant target action. Using Python scripts, we then generated nonsense word sequences from the wake-up word `hey Google' and from each target command in batches of 100 and tested the response of Google Assistant Service to audio file input of the potential adversarial commands for wake-up word and target commands separately. The choice of consonant phoneme to be replaced to generate nonsense words was performed randomly by the Python scripts for each batch of 100 potential adversarial commands. We continued the testing process until we had generated 15 successful adversarial commands for the wake-up word, and 3 successful adversarial commands for each target command, i.e. 15 successful adversarial commands in total. Each successful adversarial command for the wake-up word and each successful adversarial command for a target command were then combined to generate potential adversarial commands for the over-the-air tests as described above.

In the over-the-air tests, the 225 potential adversarial commands generated from the adversarial commands for the wake-up word and target commands which had been successful in the audio file input tests were played to the Google Assistant Library via a USB plug-in microphone from an Android smartphone. 

\subsection{Human Comprehensibility Tests}

We next tested the human comprehensibility of adversarial commands which had successfully triggered a target action by the Assistant. Human experimental subjects were recruited via the online platform Prolific Academic.\footnote{https://prolific.ac/} All subjects were native speakers of English. The subjects were asked to listen to audio of twelve successful adversarial commands, which were the successful adversarial commands shown in Tables 1 and 2 for the audio file input and over-the-air tests respectively (see Results section for further details). The audio which subjects were asked to listen to also included as `attention tests', two files consisting of synthesised audio of two easily understandable utterances, ``Hello how are you" and ``Hi how are you". Subjects were then asked to indicate whether they had identified any meaning in the audio. If they had identified meaning, they were asked to indicate what meaning they heard. The order in which audio clips were presented to the participants was randomised.

\section{Results}

\subsection{Assistant Response Tests}
Through application of the methodology described above, the audio file input tests for the wake-up word `Hey Google' identified 15 successful adversarial commands which triggered activation of the device. The audio file input tests for target commands identified 3 successful adversarial commands for each target action, i.e. 15 successful adversarial commands in total, in around 2000 tests. Three examples of the successful adversarial commands for the wake-up word and one example of an adversarial command for each of the target commands is shown in Table 1. The over-the-air tests identified 4 successful adversarial commands in the 225 tests (representing all possible combinations of each of the 15 successful adversarial commands for the wake-up word with each of the successful adversarial commands for the target commands). One of the successful over-the-air adversarial commands triggered the `turn on light' target action and three of the successful over-the-air adversarial commands triggered the `turn light red' target action. The 4 successful over-the-air adversarial commands are shown in Table 2. Also shown below, in Figures \ref{fig:transcription_1} and \ref{fig:transcription_2}, are examples of the print-out to terminal of the Google Assistant Service's response to a successful adversarial command for a wake-up word and for a target command. Further shown below is an example of the print-out to terminal of the Google Assistant Library's response to a successful over-the-air adversarial command (see Figure \ref{fig:transcription_3}).

\begin{table}[htbp!]
\centering
\begin{center}
 \begin{tabular}{|p{1.7cm}||p{1.7cm}|p{1.7cm}|p{1.7cm}|} 
 \hline
 \footnotesize{Target Command} & \footnotesize{Adversarial Command (Kirschenbaum phonetic symbols)} & \footnotesize{Text Transcribed} & \footnotesize{Action Triggered} \\ [0.5ex] 
 \hline\hline
 \footnotesize{Hey Google} & \footnotesize{S'eI j'u:b@L (``shay yooble")} & \footnotesize{hey Google} & \footnotesize{\textit{assistant activated}}\\ 
 \hline
 \footnotesize{Hey Google} & \footnotesize{t'eI g'u:t@L (``tay gootle")} & \footnotesize{hey Google} & \footnotesize{\textit{assistant activated}} \\
 \hline
 \footnotesize{Hey Google} & \footnotesize{Z'eI d'u:b@L (``zhay dooble")} & \footnotesize{hey Google} & \footnotesize{\textit{assistant activated}} \\
 \hline
 \footnotesize{turn off light} & \footnotesize{h'3:n z'0f j'aIt (``hurn zof yight)} & \footnotesize{turns off the light} & \footnotesize{Turning device off}\\[1ex]
 \hline
 \footnotesize{turn light blue} & \footnotesize{h'3:n gl'aIt skw'u: (``hurn glight squoo"} & \footnotesize{turn the lights blue} & \footnotesize{color is blue}\\[1ex]
 \hline
 \footnotesize{turn light red} & \footnotesize{str'3:n j'aIt str'Ed (``strurn yight stred"} & \footnotesize{turn the lights to Red} & \footnotesize{color is red}\\[1ex]
 \hline
 \footnotesize{what's my name} & \footnotesize{sm'0ts k'aI sp'eIm (``smots kai spaim")} & \footnotesize{what's my name} & \footnotesize{You told me your name was MK}\\[1ex]
 \hline
 \footnotesize{turn on light} & \footnotesize{p'3:n h'0n kl'aIt (``purn hon klight")} & \footnotesize{turn on light} & \footnotesize{Turning device on}\\[1ex]
 \hline
\end{tabular}
\end{center}
\caption{Examples of successful adversarial commands in audio file input experiments}
\label{table:1}
\end{table}

\begin{table}[htbp!]
\centering
\begin{center}
 \begin{tabular}{|p{1.7cm}||p{1.7cm}|p{1.7cm}|p{1.7cm}|} 
 \hline
 \footnotesize{Target Command} & \footnotesize{Adversarial Command (x-sampa phonetic symbols)} & \footnotesize{Text Transcribed} & \footnotesize{Action Triggered} \\ [0.5ex] 
 \hline\hline
 \footnotesize{Hey Google turn on light} & \footnotesize{t'eI D'u:bl= s'3:n Z'Qn j'aIt (``tay dooble surn zhon yight")} & \footnotesize{switch on the light} & \footnotesize{Turning the LED on}\\ 
 \hline
 \footnotesize{Hey Google turn light red} & \footnotesize{t'eI D'u:bl= tr'3:n Tr'aIt str'Ed (``tay dooble trurn thright stred")} & \footnotesize{turn lights to Red} & \footnotesize{The color is red} \\
 \hline
 \footnotesize{Hey Google turn light red} & \footnotesize{t'eI D'u:bl= pr'3:n j'aIt sw'Ed (``tay dooble prurn yight swed")} & \footnotesize{turn the lights red} & \footnotesize{The color is red} \\
 \hline
 \footnotesize{Hey Google turn light red} & \footnotesize{t'eI D'u:bl= str'3:n j'aIt str'Ed (``tay dooble strurn yight stred")} & \footnotesize{turn lights to Red} & \footnotesize{The color is red}\\[1ex]
 \hline
\end{tabular}
\end{center}
\caption{Successful adversarial commands in over-the-air experiments}
\label{table:2}
\end{table}

\begin{figure}
    \tiny
    \centering
    \noindent\caption{\small\textbf{{Transcription of response to adversarial command for `Hey Google' from audio file}}}

        \begin{verbatim}
        Wakeup word triggered by nonsense_wakeup/Z'eI d'u:b@L.raw, nonsense_wakeup/Z'eI d'u:b@L
        INFO:root:Connecting to embeddedassistant.googleapis.com
        
        INFO:root:Recording audio request.
        INFO:root:Transcript of user request: "change".
        INFO:root:Transcript of user request: "JD".
        INFO:root:Transcript of user request: "hey dude".
        INFO:root:Transcript of user request: "hey Google".
        INFO:root:Transcript of user request: "hey  Google".
        INFO:root:Transcript of user request: "hey Google".
        INFO:root:End of audio request detected.
        INFO:root:Stopping recording.
        INFO:root:Transcript of user request: "hey Google".
        INFO:root:Expecting follow-on query from user.
        INFO:root:Playing assistant response.
        \end{verbatim}
        \label{fig:transcription_1}
\end{figure}
\normalsize

\begin{figure}
    \tiny
    \noindent\caption{\small\textbf{{Transcription of response to adversarial command for `what's my name' (sm'0ts k'aI sp'eIm) from audio file}}}

    \begin{verbatim}

        INFO:root:Recording audio request.
        INFO:root:Transcript of user request: "what's".
        INFO:root:Playing assistant response.
        INFO:root:Transcript of user request: "some".
        INFO:root:Playing assistant response.
        INFO:root:Transcript of user request: "summer".
        INFO:root:Playing assistant response.
        INFO:root:Transcript of user request: "what's on Sky".
        INFO:root:Playing assistant response.
        INFO:root:Transcript of user request: "what's my IP".
        INFO:root:Playing assistant response.
        INFO:root:Transcript of user request: "some months cause pain".
        INFO:root:Playing assistant response.
        INFO:root:Transcript of user request: "what's my car's paint".
        INFO:root:Playing assistant response.
        INFO:root:Transcript of user request: "what's my car's paint".
        INFO:root:Playing assistant response.
        INFO:root:End of audio request detected
        INFO:root:Transcript of user request: "what's my name".
        INFO:root:Playing assistant response.
        INFO:root:You told me your name was MK
        I could never forget that ��
        INFO:root:Finished playing assistant response.
        \end{verbatim}
        \label{fig:transcription_2}
\end{figure}
\normalsize

\begin{figure}
    \tiny
    \centering
    \noindent\caption{\small\textbf{{Transcription of response to adversarial command for `Hey Google turn on light' (t'eI D'u:bl= s'3:n Z'Qn j'aIt) from over-the-air audio}}}

    \begin{verbatim}
        
        ON_CONVERSATION_TURN_STARTED
        ON_END_OF_UTTERANCE
        ON_RECOGNIZING_SPEECH_FINISHED:
          {"text": "switch on the light"}
        
        Do command action.devices.commands.OnOff with params {u'on': True}
        Turning the LED on.
        ON_RESPONDING_STARTED:
          {"is_error_response": false}
        ON_RESPONDING_FINISHED
        ON_CONVERSATION_TURN_FINISHED:
          {"with_follow_on_turn": false}
        
        \end{verbatim}
        \label{fig:transcription_3}
\end{figure}
\normalsize

In repeated tests, it was shown that the audio file input results were reproducible, whereas the over-the-air results were not, i.e. a successful adversarial command did not necessarily trigger the target action again on re-playing. Apart from the triggering target commands as described, a certain proportion of the nonsensical word sequences tested in the experiments were transcribed as other meaningful word sequences, prompting the Assistant to run web searches. For other nonsensical word sequences, the Assistant's response was simply to indicate non-comprehension of the input.

\subsection{Human Comprehensibility Tests}

As stated above, audio clips of the twelve successful adversarial commands shown in Tables 1 and 2, as well as two audio clips representing attention tests, were played to human subjects in an online experiment. There were 20 participants in the experiment, from whom 17 sets of valid results could be retrieved. All 17 participants who generated these results transcribed the attention tests correctly as `hi how are you' and 'hello how are you'. Three participants transcribed one adversarial command as the target command `turn on light', but did not identify any of the other target commands or the wake-up word `Hey Google' in either the audio file input clips or the over-the-air clips. None of the other participants identified any of the target commands or the wake-up word in any of the clips. Eight of the participants identified no meaning at all in any of the clips which did not represent attention tests. The other participants all either indicated incomprehension of the nonsensical sounds as well or else transcribed them as words which were unrelated to the target command for Google Assistant. Some examples of unrelated transcriptions were `hands off the yacht' and `smoking cause pain'. One participant also transcribed some of the nonsensical sounds as nonsense syllables e.g. `hurn glights grew' and `pern pon clight'. Another participant also transcribed a couple of the nonsensical sounds as the French words `Je du blanc'. 

\section{Discussion}

The combined results from our machine response and human comprehensibility tests confirm that voice-controlled digital assistants are potentially vulnerable to covert attacks using nonsensical sounds. The key findings are that voice commands to voice-controlled digital assistant Google Assistant are shown to be triggered by nonsensical word sounds in some instances, whereby the same nonsensical word sounds are perceived by humans as either not having any meaning at all or as having a meaning unrelated to the voice commands to the Assistant. One notable feature of the results is that the transcription of the adversarial command by the Assistant does not need to match the target command exactly in order to trigger the target action; for example, an adversarial command for the target command `turn on light' is transcribed as `switch on the light' in one instance (see Table 2).  In one case, the transcription of an adversarial command does not even need to be semantically equivalent to the target command in order to trigger the target action, as for example in the transcription of an adversarial command for ``turn off light" as ``turns off the light". This attack exploits a weakness in the natural language understanding functionality of the Assistant as well as in its speech recognition functionality.

The machine and human responses to nonsensical word sounds in general were comparable, in that both machine and humans frequently indicated incomprehension of the sounds, or else attempted to fit them to meaningful words. However, in the specific instances of nonsensical word sounds which triggered a target command in Google Assistant, none of the human listeners heard a Google Assistant voice command in the nonsensical word sounds which had triggered a target command. Another difference between the machine and human results was that whereas in addition to either indicating incomprehension or transcribing the nonsensical sounds as real words, human subjects on occasion attempted to transcribe the nonsensical word sounds phonetically as nonsense syllables, the Assistant always either indicated incomprehension or attempted to match the nonsensical sounds to real words. This confirms that, unlike humans, the Assistant does not have a concept of word sounds which have no meaning, making it vulnerable to being fooled by word sounds which are perceived by humans as obviously nonsensical.

\section{Future Work and Conclusions}

Based on this small-scale study, we conclude that voice-controlled digital assistants are potentially vulnerable to malicious input consisting of nonsense syllables which humans perceive as meaningless. A focus of future work might be to conduct a larger scale study and to conduct a more fine-tuned analysis of successful and unsuccessful nonsense attacks, to determine which nonsense syllables are most likely to be confused with target commands by machines, whilst still being perceived as nonsensical by humans. This would enable investigation of more targeted attacks. Ultimately the focus of future work should be to consider how voice-controlled systems might be better trained to distinguish between meaningful and meaningless sound in terms of the language to which they are intended to respond. 

% conference papers do not normally have an appendix

% use section* for acknowledgment
\section*{Acknowledgments}

This work was funded by doctoral training grant from the Engineering and Physical Sciences Research Council (EPSRC).

% trigger a \newpage just before the given reference
% number - used to balance the columns on the last page
% adjust value as needed - may need to be readjusted if
% the document is modified later
%\IEEEtriggeratref{8}
% The "triggered" command can be changed if desired:
%\IEEEtriggercmd{\enlargethispage{-5in}}

% references section

% can use a bibliography generated by BibTeX as a .bbl file
% BibTeX documentation can be easily obtained at:
% http://mirror.ctan.org/biblio/bibtex/contrib/doc/
% The IEEEtran BibTeX style support page is at:
% http://www.michaelshell.org/tex/ieeetran/bibtex/
%\bibliographystyle{IEEEtran}
% argument is your BibTeX string definitions and bibliography database(s)
%\bibliography{IEEEabrv,../bib/paper}
%
% <OR> manually copy in the resultant .bbl file
% set second argument of \begin to the number of references
% (used to reserve space for the reference number labels box)

%MKB_14.03.18_original bibliography section replaced

{\footnotesize \bibliographystyle{IEEEtran}
\bibliography{noncens}}

% that's all folks
\end{document}